\documentclass[ALICE,manyauthors]{cernphprep}

\usepackage[comma,square,numbers,sort&compress]{natbib}
\usepackage{hyperref}
\usepackage{lineno}
\usepackage{subfigure}
\usepackage[section]{placeins}
\usepackage {multicol}
\usepackage {multirow}
\usepackage {units}
\usepackage {array}

\usepackage[T1]{fontenc}

\newcommand{\pt}           {\ensuremath{p_\mathrm{T}}}

\newcommand{\pythiam}{\ensuremath{\mathrm{PYTHIA}\,8\,\mathrm{Tune\,4C}}}
\newcommand{\pythiashoving}{{\ensuremath{\mathrm{PYTHIA}\,8~\mathrm{String}~}}\ensuremath{\mathrm{Shoving}}}
\newcommand{\epos}{{\ensuremath{\mathrm{EPOS}\,~\mathrm{LHC}}}}
\newcommand{\pttrig}{\ensuremath{p_\mathrm{T,\,trig}}}
\newcommand{\ptassoc}{\ensuremath{p_\mathrm{T,\,assoc}}}
\newcommand{\ptjet}{\ensuremath{p_\mathrm{T,\,jet}^\mathrm{ch}}}
\newcommand{\ptlead}{\ensuremath{p_\mathrm{T,\,LP}}}
\newcommand{\pttrigassoc}{\ensuremath{p_\mathrm{T,\,trig\,(assoc)}}}

\begin{document}

\begin{titlepage}
\PHyear{2021}       
\PHnumber{006}      
\PHdate{6 January}  

\title{Long- and short-range correlations and their event-scale dependence in high-multiplicity pp collisions at $\boldsymbol{\sqrt{{\textit s}}}=13$~TeV}
\ShortTitle{Long-range correlations in pp collisions at $\sqrt{s} = 13$~TeV}   

\Collaboration{ALICE Collaboration\thanks{See Appendix~\ref{app:collab} for the list of collaboration members}}
\ShortAuthor{ALICE Collaboration} 

\begin{abstract}
Two-particle angular correlations are measured in high-multiplicity proton--proton collisions at $\sqrt{s} =13$~TeV by the ALICE Collaboration. The yields of particle pairs at short-($\Delta\eta$ $\sim$ 0) and long-range ($1.6 < |\Delta\eta| < 1.8$) in pseudorapidity are extracted on the near-side ($\Delta\varphi$ $\sim$ 0).
They are reported as a function of transverse momentum ($p_{\mathrm T}$) in the range 1~$<p_{\mathrm T}<$~4 GeV/$c$.
Furthermore, the event-scale dependence is studied for the first time by requiring the presence of high-$p_{\rm T}$ leading particles or jets for varying $p_{\rm T}$ thresholds. 
The results demonstrate that the long-range ``ridge'' yield, possibly related to the collective behavior of the system, is present in events with high-$p_{\mathrm T}$ processes as well. The magnitudes of the short- and long-range yields are found to grow with the event scale. 
The results are compared to EPOS~LHC and PYTHIA~8 calculations, with and without string-shoving interactions. It is found that while both models describe the qualitative trends in the data, calculations from EPOS~LHC show a better quantitative agreement for the $p_{\rm T}$ dependency, while overestimating the event-scale dependency.
\end{abstract}

\end{titlepage}

\setcounter{page}{2}


\section{Introduction}
\label{sec:intro}

In high-energy nucleus--nucleus collisions at RHIC~\cite{Adams:2005dq,Adcox:2004mh,Arsene:2004fa,Back:2004je} and LHC~\cite{Abelev:2012di, Abelev:2014pua, ATLAS:2011ah}, significant correlations are observed between particles emitted over a wide pseudorapidity range. The origin of these observations are collective effects, which are related to the formation of a strongly interacting quark--gluon plasma (QGP), which exhibits hydrodynamic behavior (see the reviews~\cite{Romatschke:2007mq,Jeon:2015dfa,Romatschke:2017ejr}). 
Recent theoretical~\cite{Niemi:2015qia,Bernhard:2016tnd,Bernhard:2019bmu} and experimental~\cite{ALICE:2016kpq,Acharya:2017gsw,Acharya:2017zfg,Acharya:2020taj} advancements have contributed significantly to the understanding of the transport properties of the QGP.
Similar long-range correlations are also observed in high-multiplicity proton--proton (pp)~\cite{Aad:2015gqa,Khachatryan:2015lva,Khachatryan:2016txc,Acharya:2019vdf}, proton--nucleus (pA)~\cite{Abelev:2012ola,Aad:2014lta,Aaboud:2016yar,Khachatryan:2016ibd}, and light nucleus--nucleus collisions~\cite{PHENIX:2018lia,Aidala:2017ajz}. The fact that these correlations extend over a large range in pseudorapidity implies that they originate from early times in these collisions and thus suggest that hydrodynamic behavior is present even in these small systems, although the volume and lifetime of the medium produced in such a collision system are expected to be small, and there are other mechanisms which can produce similar flow-like signals~\cite{Busza:2018rrf,Nagle:2018nvi}.

Measurements of two-particle angular correlations provide information on many physical effects, including collectivity, hadronization, fragmentation, and femtoscopic effects~\cite{Lisa:2005dd}, and are typically quantified as a function of $\Delta\eta$, the relative pseudorapidity, and $\Delta\varphi$, the separation in azimuthal angle, of particle pairs. The long-range structure of two-particle angular correlations is well suited to analyze collective effects, since it is not created by resonance decays nor fragmentation of high-momentum partons. A typical source of long-range correlations in Monte Carlo pp generators is the momentum conservation.
The enhancement in the yield of two-particle correlations at small $\Delta\varphi$ that extends over a large $\Delta\eta$ is dubbed ``ridge'' due to its characteristic shape in the $\Delta\eta$--$\Delta\varphi$ plane.
The shape of these $\Delta\varphi$ correlations can be studied via a Fourier decomposition~\cite{Poskanzer:1998yz,Voloshin:2008dg}. The second and third order terms are the dominant harmonic coefficients. In heavy-ion collisions, harmonic coefficients can be related to the collision geometry and density fluctuations of the colliding nuclei~\cite{Alver:2010gr,Alver:2010dn,ALICE:2011ab} and to transport properties of the QGP in relativistic viscous hydrodynamic models~\cite{Gale:2012rq,Niemi:2015qia,Shen:2014vra,Bernhard:2016tnd,Bernhard:2019bmu}.

The ridge structures in high-multiplicity pp and p--Pb events have been attributed to initial-state or final-state effects. Initial-state effects, usually attributed to gluon saturation~\cite{Dusling:2012cg,Bzdak:2013zma}, can form long-range
correlations along the longitudinal direction. The final-state effects might be parton-induced interactions~\cite{Arbuzov:2011yr} or collective phenomena due to hydrodynamic behavior of the produced matter arising in a high-density system possibly formed in these collisions~\cite{Weller:2017tsr,Zhao:2017rgg}. 
Hybrid models implementing both effects are generally used in hydrodynamic simulations~\cite{Greif:2017bnr,Mantysaari:2017cni}. EPOS LHC describes collectivity in small systems with a parameterized hydrodynamic evolution of the high-energy density region, so called ``core'', formed by many color string fields~\cite{Pierog:2013ria}.
The proton shape and its fluctuations are also important to model small systems~\cite{Mantysaari:2017cni}.
To understand the influence of initial- or final-state effects, and to possibly disentangle the two, a quantitative description of the measurements in small systems~\cite{Schenke:2019pmk,Schenke:2020mbo} needs to account for details of the initial state.
Systematic studies of these correlation effects from small to large systems are being performed, both experimentally~\cite{Acharya:2019vdf} and theoretically~\cite{Schenke:2020mbo}.
However, the quantitative description of the full set of experimental data has not been achieved yet.
A summary of various explanations for the observed correlations in small systems is given in~\cite{Strickland:2018exs,Loizides:2016tew,Nagle:2018nvi,Loizides:2016tew}.

Besides the hybrid models mentioned above, alternative approaches were developed to describe collectivity in small systems. A microscopic model for collectivity was implemented in the PYTHIA~8 event generator, which is based on interacting strings (string shoving) and is called the “string shoving model”~\cite{Bierlich:2017vhg}. In this model, strings repel each other in the transverse direction, which results in microscopic transverse pressure and, consequently, in long-range correlations. PYTHIA~8 with string shoving can qualitatively reproduce the near-side ($\Delta\varphi\sim0$) ridge yield measured by the CMS Collaboration~\cite{Khachatryan:2016txc}. This challenges the hydrodynamic picture and predicts modifications of the jet fragmentation properties~\cite{Bierlich:2019ixq}.

It is expected that final-state interactions affect also produced jets if they are the source of collectivity in small systems. Proving the presence of jet quenching~\cite{Gyulassy:1990ye,Wang:1991xy} would be another crucial evidence of the existence of a high-density strongly-interacting system, possibly a QGP, in high-multiplicity pp collisions. However, there is no evidence observed so far for the jet quenching effect in high-multiplicity pp and p--Pb collisions~\cite{Khachatryan:2016odn,Adam:2016jfp,Adam:2016dau,Acharya:2017okq}. Jet fragmentation can be studied in two-particle angular correlations in short-range correlations around $(\Delta\eta$, $\Delta\varphi)=(0,0)$~\cite{Adam:2016tsv}.  

To further investigate the interplay of jet production and collective effects in small systems, long- and short-range correlations are studied simultaneously in high-multiplicity pp collisions at $\sqrt{s} =13$ TeV using the ALICE LHC Run 2 data collected with the high-multiplicity event trigger in 2016--2018. In this article, the near-side per-trigger yield at large pseudorapidity separation is presented as a function of transverse momentum. The results are compared with previous measurements by the CMS Collaboration~\cite{Khachatryan:2015lva}. In addition, the ridge yield and near-side jet-like correlations with the event-scale selection are reported. The event-scale selection is done by requiring a minimum transverse momentum of the leading particle or the reconstructed jet at midrapidity, which is expected to bias the impact parameter of pp collisions to be smaller on average~\cite{Sjostrand:1986ep,Frankfurt:2010ea}. At the same time, the transverse momentum of the leading particle or the reconstructed jet is a measure of the momentum transfer in the hard parton scattering~\cite{Chatrchyan:2012tt,Chatrchyan:2011id}. The event-scale dependence of the second-flow harmonic $v_{2}$ has previously been studied in pp collisions with and without a tagged $Z$ boson, where little or no dependence was observed~\cite{Aaboud:2019mcw}.

The experimental setup and analysis method are described in Sec.~\ref{sec:experiment} and~\ref{sec:ana}, respectively. The sources of systematic uncertainties are discussed in Sec.~\ref{sec:uncertainties}. The results and comparisons with model calculations of the measurements are presented in Sec.~\ref{sec:results}. Finally, results are summarized in~Sec.~\ref{sec:summary}.

\section{Experimental setup}
\label{sec:experiment}

The analysis is carried out with data samples of pp collisions at $\sqrt{s} = 13$~TeV collected from 2016 to 2018 during the LHC Run 2 period. The full description of the ALICE detector and its performance in the LHC Run 2 can be found in~\cite{Aamodt:2008zz,Abelev:2014ffa}. The present analysis utilizes the V0~\cite{Abbas:2013taa}, the Inner Tracking System (ITS)~\cite{aliceITS}, and the Time Projection Chamber (TPC)~\cite{aliceTPC} detectors.

The V0 detector consists of two stations placed on both sides of the interaction point, V0A and V0C, each made of 32 plastic scintillator tiles, covering the full azimuthal angle within the pseudorapidity intervals $2.8 < \eta < 5.1$ and $-3.7 < \eta < -1.7$, respectively. The V0 is used to provide a minimum bias (MB) and a high-multiplicity (HM) trigger. The minimum bias trigger is obtained by a time coincidence of V0A and V0C signals. The charged particle multiplicity selection is done on the sum of the V0A and V0C signals, which is denoted as V0M. The high-multiplicity trigger requires that the V0M signal exceeds 5 times the mean value measured in minimum bias collisions, selecting the 0.1\% of MB events that have the largest V0 multiplicity. The analyzed data samples of minimum bias and high-multiplicity pp events at $\sqrt{s}=$~13 TeV correspond to integrated luminosities of 19 nb$^{-1}$ and 11 pb$^{-1}$, respectively~\cite{ALICE-PUBLIC-2016-002}.

The primary vertex position is reconstructed from the measured signals in the Silicon Pixel Detector (SPD), which forms the innermost two layers of the ITS. Reconstructed primary vertices of selected events are required to be located within 8~cm from the center of the detector along the beam direction. The probability of pileup events is about 0.6\% in MB events. Pileup events can be resolved and are rejected if the longitudinal displacement of their primary vertices is larger than 0.8 cm.

Charged-particle tracks are reconstructed by the ITS and TPC, which are operated in a uniform solenoidal magnetic field of 0.5~T along the beam direction. The ITS is a silicon tracker with six layers of silicon sensors where the SPD~\cite{Santoro2009:ALICESPD} comprises the two innermost layers, the next two layers called the Silicon Drift Detector (SDD), and the outermost layers named the Silicon Strip Detector (SSD). The ITS and TPC, covering the full azimuthal range, have acceptances up to $|\eta| < 1.4$ and 0.9, respectively, for detection of charged particles emitted within 8 cm from the primary vertex position ($z_\mathrm{vtx}$) along the beam direction. The tracking of charged particles is done with the combined information of the ITS and TPC that enables the reconstruction of tracks down to 0.15~GeV/$c$, where the efficiency is about 65\%. The efficiency reaches 80\% for intermediate $\pt$, 1 to 5~GeV/$c$. The $\pt$ resolution is around 1\% for primary tracks with $\pt<$~1~GeV/$c$, and linearly increases up to 6\% at $\pt \sim$ 40~GeV/$c$~\cite{Contin_2012:ITSPTRES}.

The charged particle selection criteria are optimized to make the efficiency uniform over the full TPC volume to mitigate the effect of small regions where some of the ITS layers are inactive. The selection consists of two track classes. Those belonging to the first class are required to have at least one hit in the SPD. Tracks from the second class do not have any SPD associated hit and their initial point is instead constrained to the primary vertex~\cite{Adam:2015ewa}.

\section{Analysis procedure}
\label{sec:ana}

The two-particle correlation function is measured as a function of the relative pseudorapidity ($\Delta\eta$) and the azimuthal angle difference ($\Delta\varphi$) between the trigger and the associated particles,
\begin{eqnarray}
\frac{1}{N_{\rm{trig}}} \frac{ \rm{d}\it{}^{2} N_{\rm{pair}} }{ \rm{d} \Delta\eta \rm{d}\Delta\varphi} = B(0, 0)\frac{S(\Delta\eta, \Delta\varphi)}{B(\Delta\eta, \Delta\varphi)}  \Big\lvert_{\pttrig,\,\ptassoc}\quad , 
\label{eq:corrfunction}
\end{eqnarray}
where $\pttrig$ and $\ptassoc$ ($\pttrig>\ptassoc$) are the transverse momenta of the trigger and associated particles, respectively, $N_\mathrm{trig}$ is the number of trigger particles, and $N_\mathrm{pair}$ is the number of trigger-associated particle pairs. The average number of pairs in the same event and in mixed events are denoted as $S(\Delta\eta, \Delta\varphi)$ and $B(\Delta\eta, \Delta\varphi)$, respectively. Normalization of $B(\Delta\eta, \Delta\varphi)$ is done with its value at $\Delta\eta$ and $\Delta\varphi = 0$, represented as $B (0,0)$. Acceptance effects are corrected by dividing $S(\Delta\eta, \Delta\varphi)$ with $B(\Delta\eta, \Delta\varphi)/B (0,0)$. The right-hand side of Eq.~(1) is corrected for the track reconstruction efficiency, which is mainly relevant for the associated particles, as a function of $\pt$ and pseudorapidity. Primary vertices of events to be mixed are required to be within the same, 2 cm wide, $z_{\rm{vtx}}$ interval~\cite{KOPYLOV1974472:evtmixing,Adam:2016tsv} for each multiplicity class. The final per-trigger yield is constructed by averaging correlation functions over these primary vertex bins.

Ridge yields at large $\Delta\eta$ are extracted for various multiplicity classes and $\pt$ intervals. The large $\Delta\eta$ range is selected as $1.6<|\Delta\eta|<1.8$, which is the range where the tracking quality -- efficiency and precision -- is the best. The ridge yield is only reported for $\pt>$~1~GeV/$c$. Below 1~GeV/$c$, the jet-like contribution to the correlation function extends into the region where the ridge yield is measured, 1.6~$<|\Delta\eta|<$~1.8. In this region, the $\Delta\varphi$ distribution, or the so-called per-trigger yield, is expressed as
\begin{eqnarray}
\frac{1}{N_{\rm{trig}}} \frac{ \rm{d}\it{}N_{\rm{pair}} }{ \rm{d}\Delta\varphi } = \int_{1.6<|\Delta \eta|<1.8} \left( \frac{1}{\it{N}_{\rm{trig}}} \frac{ \rm{d}\it{}^{2} N_{\rm{pair}} }{ \rm{d}\Delta\eta d\Delta\varphi} \right) \dfrac{1}{\delta_{\Delta\eta}} \rm{d}\Delta \eta - C_{\rm{ZYAM}}\quad ,
\end{eqnarray}
where $\delta_{\Delta\eta}=$~0.4 is the normalization factor to get the per-trigger yield per unit of pseudorapidity. 

The baseline of the correlation is subtracted by means of the Zero-Yield-At-Minimum (ZYAM) procedure~\cite{Ajitanand:2005jj}. The minimum yield $(C_{\rm{ZYAM}})$ at $\Delta\varphi=\Delta\varphi_{\rm{min}}$ in the $\Delta\varphi$ projection (note that the value of $\Delta\varphi_{\rm{min}}$ can be different in data and in models) is obtained from a fit function, which fits the data with a Fourier series up to the third harmonic. By construction, the yield at $\Delta\varphi_{\rm{min}}$ is zero after subtracting $C_{\rm{ZYAM}}$ from the $\Delta\varphi$ projection. The ridge yield ($Y^{\rm{ridge}}$) is obtained by integrating the near-side peak of the $\Delta\varphi$ projection over $|\Delta\varphi|<|\Delta\varphi_{\rm{min}}|$ after the ZYAM procedure,
\begin{eqnarray}
Y^{\rm{ridge}} = \int_{|\Delta \varphi| < |\Delta\varphi_{\rm{min}}| } \frac{1}{\it{N}_{\rm{trig}}} \frac{ \rm{d}\it{}N_{\rm{pair}} }{ \rm{d}\Delta\varphi }  \rm{d} \Delta\varphi .
\end{eqnarray}

The ridge yield is further studied in events having a hard jet or a high-$\pt$ leading particle in the midrapidity region. This event scale is set by requiring a minimum $\pt$ of the leading track ($\ptlead$) or the reconstructed jet ($\ptjet$) at midrapidity. The leading track is selected within $|\eta|<0.9$ and the full azimuthal angle. Jets are reconstructed with charged particles only (track-based jets) with the anti-$k_{\rm{T}}$ algorithm~\cite{Cacciari:2008gp,Cacciari:2011ma} and the resolution parameter $R=$~0.4. The recombination scheme used in this article is the $\pt$ scheme. Jets are selected in $|\eta_\mathrm{jet}|<0.4$ and in the full azimuthal angle. The $\pt$ of jets $\ptjet$ is corrected for the underlying event density that is measured using the $k_{\rm{T}}$ algorithm with $R=$~0.2~\cite{Acharya:2018eat}. 

To quantify the variation of the near-side jet-like peak with event-scale selections with a minimum $p_\mathrm{T,\,LP}$ or $\ptjet$, the near-side jet-like peak yield is extracted from the near-side $\Delta\eta$ correlations. The near-side is defined as $|\Delta\varphi|<$~1.28, where the correlation function is projected on the $\Delta\eta$ axis. The projection range, 1.28, is chosen to fully cover $\Delta\varphi_{\rm{min}}$. The near-side $\Delta\eta$ correlations are then constructed as
\begin{eqnarray}
\frac{1}{N_{\rm{trig}}} \frac{ \rm{d}\it{}N_{\rm{pair}} }{ \rm{d}\Delta\eta } = \int_{|\Delta \varphi|<1.28} \left( \frac{1}{\it{N}_{\rm{trig}}} \frac{ \rm{d}\it{}^{2} N_{\rm{pair}} }{ \rm{d}\Delta\eta d\Delta\varphi} \right) \dfrac{1}{\delta_{\Delta\varphi}} \rm{d} \Delta \varphi - \it{D}_{\rm{ZYAM}} \quad,
\end{eqnarray}
where $\delta_{\Delta\varphi}=$~2.56 is the normalization factor to get per-trigger yield per unit of azimuthal angle.
The minimum yield $(D_{\rm{ZYAM}})$ of the $\Delta\eta$ correlations is found within $|\Delta\eta|<$~1.6 and used for the subtraction from  the $\Delta\eta$ correlations, which results in zero-yield at the minimum. The near-side jet-like peak yield ($Y^{\mathrm{near}}$) is measured by integrating the $\Delta\eta$ correlations over $|\Delta\eta|<$~1.6,
\begin{eqnarray}
Y^{\rm{near}} = \int_{|\Delta \eta|<1.6} \left( \frac{1}{\it{N}_{\rm{trig}}} \frac{ \rm{d}\it{}N_{\rm{pair}} }{ \rm{d}\Delta\eta } \right) \rm{d} \Delta\eta
\end{eqnarray}

\section{Systematic uncertainties of the measured yields}
\label{sec:uncertainties}

The systematic uncertainties of $Y^{\rm{ridge}}$ and $Y^{\rm{near}}$ are estimated by varying the analysis selection criteria and corrections and are summarized in Tab.~\ref{tab:syst}.

\begin{table}[h!]
\caption{The relative systematic uncertainty of $Y^{\rm{ridge}}$ and $Y^{\rm{near}}$. Numbers given in ranges correspond to minimum and maximum uncertainties.}
\centering
\begin{tabular}{c|cc}
\hline 
\multirow{2}{*}{Sources}  & \multicolumn{2}{c}{Systematic uncertainty (\%)} \\\cline{2-3} 
         & $Y^{\rm{ridge}}$ & $Y^{\rm{near}}$ \\ \hline 
Pileup rejection    	& $\pm$0.8--3.9    &$\pm$0.2--2.2	\\ 
Primary vertex	        & $\pm$0.5--2.4	   &$\pm$1.1	\\ 
Tracking		        & $\pm$2.0--4.0    &$\pm$1.5--3.4	\\ 
ZYAM		        	& $\pm$2.1--5.1	   &$\pm$2.2--4.8	\\ 
Event mixing	    	& $\pm$1.0--4.4	   &$\pm$0.5--1.7	\\ 
Efficiency correction	& $\pm$2.5 	    &$\pm$3.1	\\  
Jet contamination   	& $-$18.8--25.9 ($\pt<$~2~GeV/$c$)	&N.A.	\\ \hline 
Total (in quadrature)			& $^{+\rm{4.9}\textrm{--}\rm{9.4}}_{-\rm{19.4}\textrm{--}\rm{21.0}}$ & $\pm$3.9--7.3 \\ 
\hline 
\end{tabular}
\label{tab:syst}
\end{table}

The systematic uncertainties are independent of the event-scale selection except for $D_{\rm{ZYAM}}$ (see below), as expected, since the multiplicity is weakly dependent on the event scale and the ALICE detector is optimized for much higher multiplicities (Pb--Pb collisions), this is in agreement with our expectations.

The uncertainty associated to the pileup rejection is estimated by measuring the changes of results with different rejection criteria from the default one. It is mainly estimated by varying the minimal number of track contributors required for reconstruction of pileup event vertices from 3 to 5. The estimated uncertainty of $Y^{\rm{ridge}}$ is 0.8-3.9\%. The corresponding uncertainty of $Y^{\rm{near}}$ is estimated to be 0.2--2.2\%.
 
Another source of systematic uncertainty is related to the selected range of the primary vertex. The accepted range is changed from $|z_\mathrm{vtx}|<$ 8 cm to $|z_\mathrm{vtx}|<$ 6 cm. The narrower primary vertex selection allows one to test acceptance effects on the measurement. The estimated uncertainty of $Y^{\rm{ridge}}$ is 0.5--2.4\%. The uncertainty for $Y^{\rm{near}}$ is estimated to be 1.1\%.

An additional source of systematic uncertainty is related to the track selection criteria. The corresponding uncertainty is estimated by employing other track selection criteria, denoted global tracks, which are optimized for particle identification. The selection criteria of the global tracks are almost identical to the hybrid tracks. Each global track is required to have at least one SPD hit. Due to inefficient parts of the SPD, the azimuthal distribution of global tracks is not uniform.
The uncertainties associated with the track selection are estimated to be 2.0--4.0\% and 1.5--3.4\% for $Y^{\rm{ridge}}$ and $Y^{\rm{near}}$, respectively.

The systematic uncertainty of $Y^{\rm{ridge}}$ resulting from the ZYAM procedure is estimated by varying the range of the fit, which is used to find the minimum, from $|\Delta\varphi|<\pi/$2 down to $|\Delta\varphi|<1.2$. The estimated uncertainty of $Y^{\rm{ridge}}$ is 2.1--5.1\%. The corresponding uncertainty on $Y^{\rm{near}}$ is estimated by varying the range from $|\Delta\eta|<$~1.6 to $|\Delta\eta|<$~1.5 and 1.7. The estimated uncertainty of $Y^{\rm{near}}$ is 2.2\% for the unbiased case and increases to 4.8\% for the largest event-scale selections. This is the only systematic uncertainty for which a significant dependence on the event scale is observed, reflecting a non-negligible dependence of the near-side magnitude and shape on the event-scale selection.

The source of systematic uncertainty is associated to the choice of the width of $z_{\rm vtx}$ bins that are used in the event mixing method. The default value of 2\,cm is changed to 1\,cm. The resulting uncertainty of $Y^{\rm{ridge}}$ is 1.0--4.4\%.
The uncertainty for $Y^{\rm{near}}$ is about 0.5--1.7\%. The uncertainty from the efficiency correction for charged particles is estimated by comparing correlation functions of true particles with correlation functions of reconstructed tracks with the efficiency correction in simulation. The estimated uncertainties are 2.5\% and 3.1\% for $Y^{\rm{ridge}}$ and $Y^{\rm{near}}$, respectively. 

In the limited $\eta$-acceptance of ALICE, the ridge structure is not flat in $\Delta\eta$ suggesting that jet-like correlations (non-flow) could contribute, implying that they would impact the ridge-yield extraction. We stress that the models used for comparisons also contain such a non-flow effect, but differences in jet-like correlations between data and MC models could influence the interpretation. To account for the related uncertainty, the variation of the yield with $\Delta\eta$ between 1.5 and 1.8, which should be an upper limit of the residual jet-like contamination, is used as a systematic uncertainty of the ridge yield. The estimated upper limit of the uncertainty is $-$25.9\% for the 1.0~$<\pt<$~1.5~GeV/$c$ range,  $-$18.8\% for the 1.5~$<\pt<$~2.0 GeV/$c$ range,  $-$18.9\% for the 1.0~$<\pt<$~2.0~GeV/$c$ range, and negligible for $\pt>$~2.0~GeV/$c$. This uncertainty is considered only for the measured ridge yields.


\section {Results}
\label{sec:results}

\subsection{Ridge yield}
\label{sec:resultunbiased}
\begin{figure}[b!]
	\centering
	\subfigure{ \includegraphics[width=0.47 \textwidth]{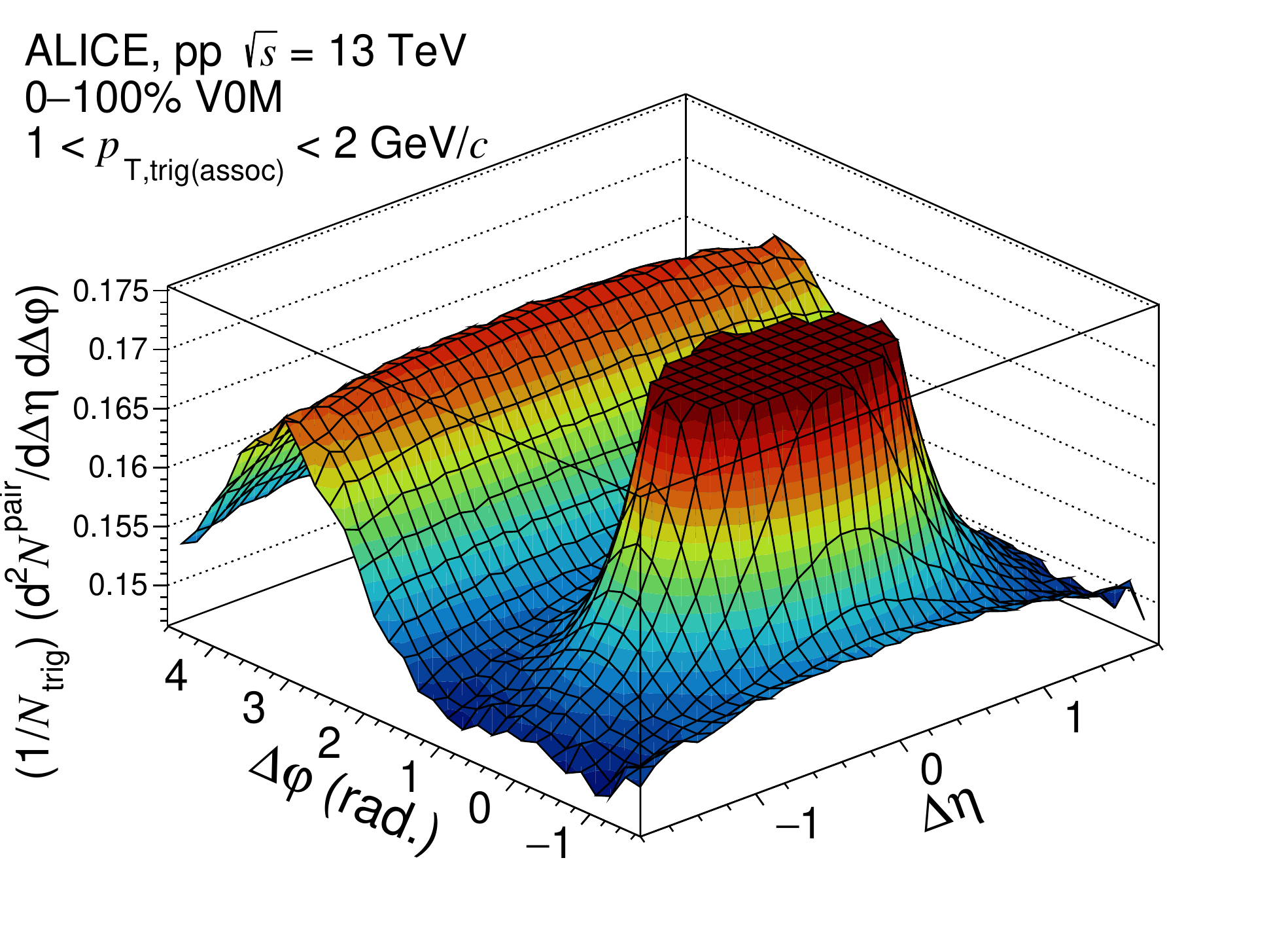} }
	\subfigure{ \includegraphics[width=0.47 \textwidth]{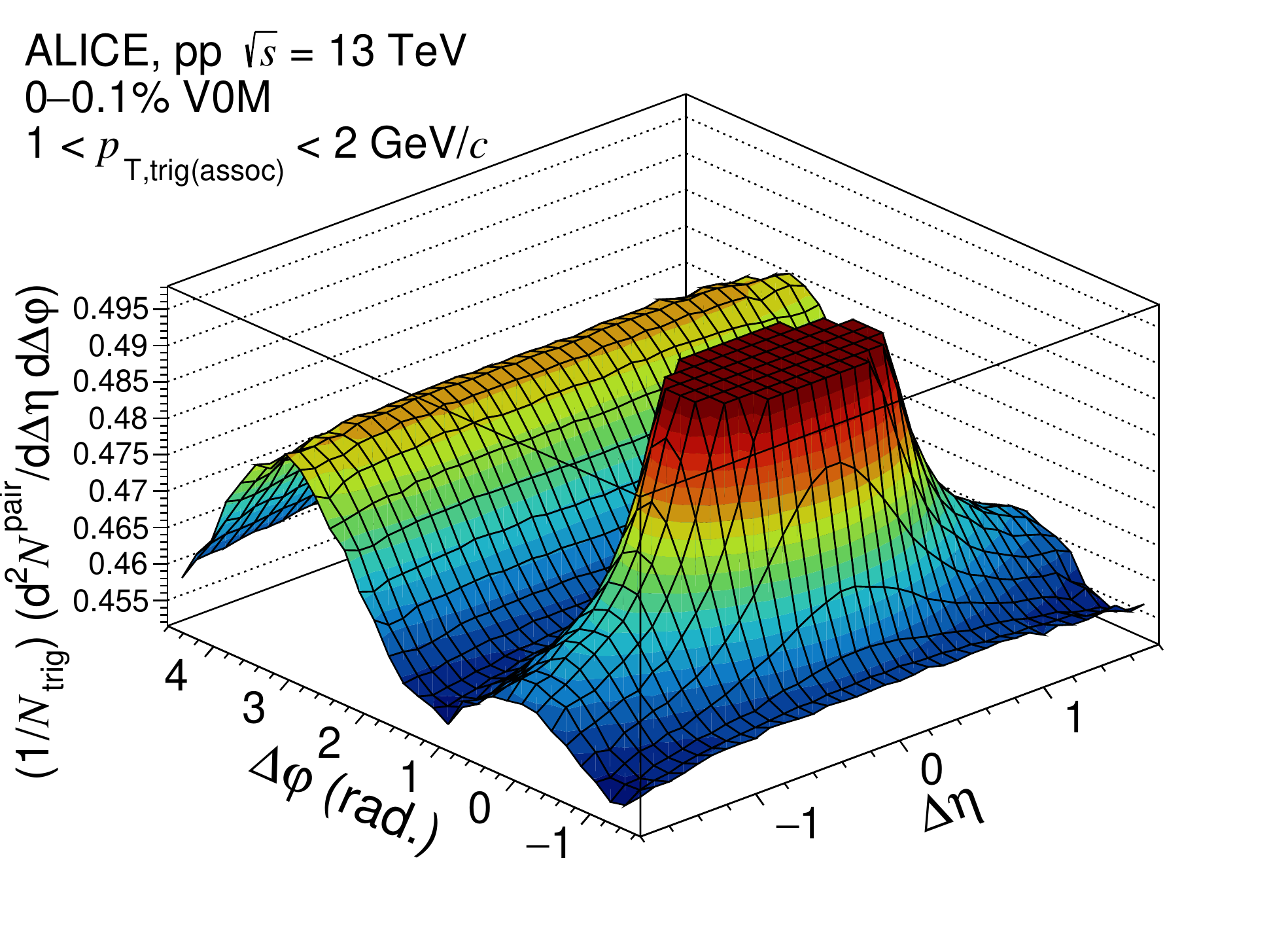} }
	\caption{ Two-particle correlation functions as functions of $\Delta\eta$ and $\Delta\varphi$ in minimum-bias events (0--100\%, left) and high-multiplicity (0--0.1\%, right). Note that the near-side jet peaks exceed the chosen range of the $z$-axis. The intervals of $\pttrig$ and $\ptassoc$ are 1~$<\it{p}_{\rm{T}}<$~2~GeV/$c$ in both cases.}
	\label{fig:PlotCorrMBHMT}
\end{figure}

Figure~\ref{fig:PlotCorrMBHMT} shows the per-trigger yield obtained from Eq.~(1) for 1~$<\pttrig~\mathrm (\ptassoc)<$~2 GeV/$c$ in pp collisions at $\sqrt{\it{s}} = $\unit{13} {\rm{}TeV} for minimum bias events (left) and high-multiplicity events (right). It is worth noting that the $z$-axes for the yield of the correlations is properly scaled in order to zoom in the ridge yield, as a result, the jet peaks are sheared off in both figures. The ridge structure is clearly observed in the high-multiplicity class while it is less significant in the minimum bias events. The away-side yield is populated mostly by back-to-back jet correlations.

\begin{figure}[h!]
	\centering
	\includegraphics[width=0.9\linewidth]{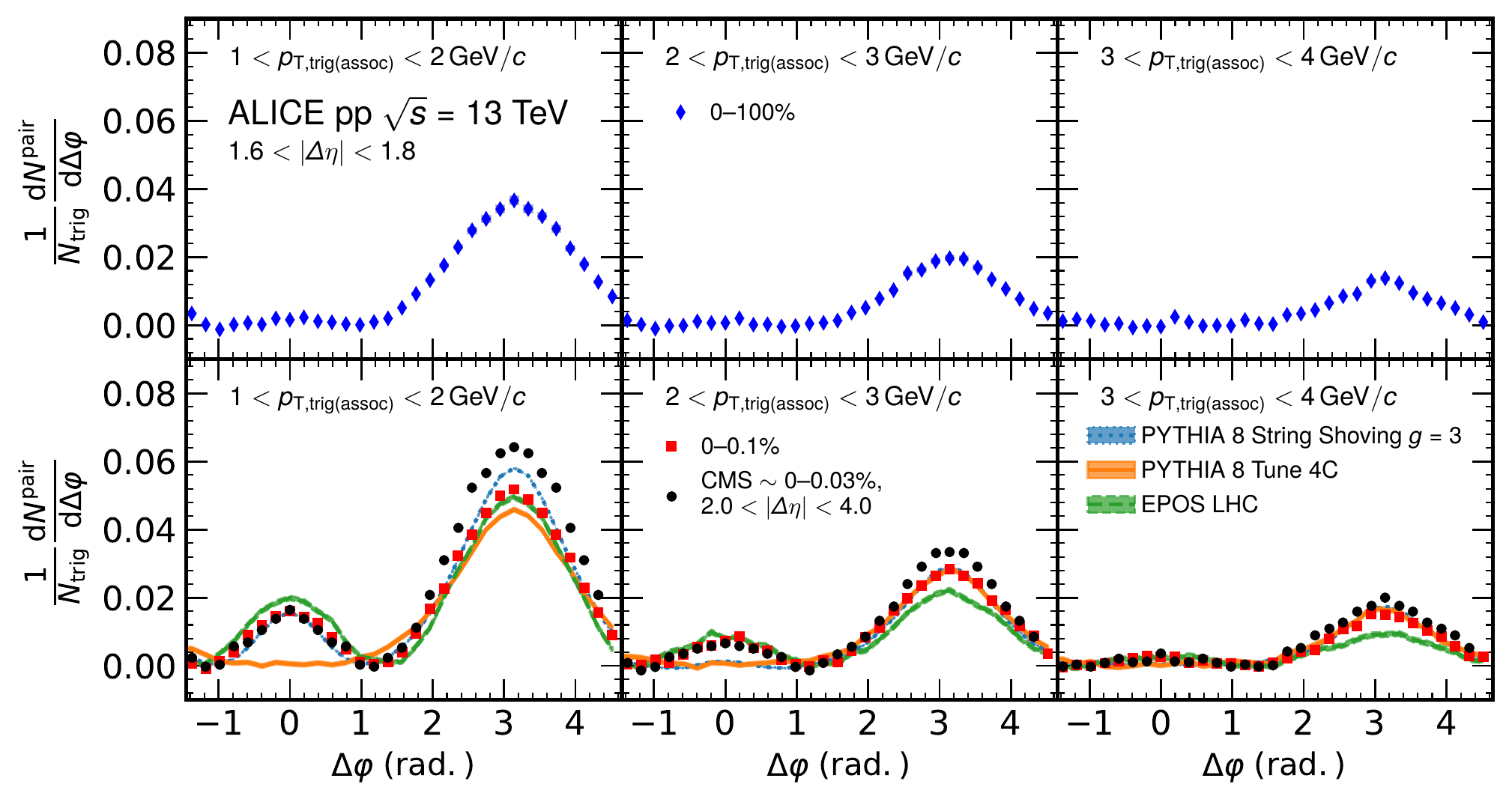}
	\caption{One-dimensional $\Delta\varphi$ distribution in the large $\Delta\eta$ projection for three transverse momentum intervals in minimum bias (upper panels) and high-multiplicity (lower panels) events after ZYAM subtraction. Transverse momentum intervals of the trigger particles and associated particles are 1~$<\pt<$~2 (left), 2~$<\pt<$~3 (middle) and, 3~$<\pt<$~4 GeV/$c$ (right), respectively. The presented model predictions were calculated using $\pythiashoving$, $\pythiam$, and $\epos$.}
	\label{fig:PlotDeltaPhi}
\end{figure}
 
Figure~\ref{fig:PlotDeltaPhi} shows $\Delta\varphi$ projections of the two-particle correlation functions obtained in the range 1.6~$<|\Delta \eta |<$~1.8 for several track $\pt$ intervals after the ZYAM subtraction (see Eq.~(2)). The results are shown for various $\it{p}_{\rm{T}}$ intervals in the minimum bias class (upper) and the high-multiplicity class (lower) down to 1~GeV/$c$ where the non-flow contamination is negligible. The near-side ($\Delta\varphi\sim 0$) ridge in the high-multiplicity class is clearly observed for $\pt<$~3~GeV/$c$ while there is no definitive signal in the minimum bias class. Within the range of analyzed particle $\pt$, the yield in the near-side ridge decreases with increasing $\pt$ in the high-multiplicity class.

The measurements in the high-multiplicity class are compared with the results published by the CMS Collaboration~\cite{Khachatryan:2015lva}. In case of the CMS measurement, the charged particle multiplicity was obtained by counting the number of particles satisfying $\pt>0.4$~GeV/$c$ in $|\eta|<$~2.4. In our analysis, event multiplicity is determined from the forward V0 detectors. The difference in multiplicity selection between ALICE (forward) and CMS (midrapidity) is studied using PYTHIA~8 simulations and it is found that the calculated multiplicity using the CMS procedure is about 20\% larger than the one from ALICE when compared in the acceptance region of the measurements reported in this article, $|\eta|<$~0.9. Near-side ridges in all transverse momentum ranges are comparable. The larger away-side yields observed in Fig.~\ref{fig:PlotDeltaPhi} for the CMS results can be attributed to the overlap in $\eta$ acceptance between the multiplicity selection and the correlation function measurement.

In Fig.~\ref{fig:PlotDeltaPhi}, the ALICE measurements are also compared with model predictions where a comparable high-multiplicity selection and $\Delta\eta$ projection range are applied. The selection of high-multiplicity events in the models is done by requiring a minimal number of charged particles emitted within the V0M detector acceptance. In case of $\pythiam$, the 0--0.1\% centrality threshold is 105 charged particles. The threshold for $\epos$ and $\pythiashoving$ are 110 and 108, respectively. The magnitude of string shoving ($g$) is set to 3.0 in this study. The statistical uncertainties due to the limited number of events for the model calculations are shown as bands in each figure. The $\pythiashoving$ provides good estimates of the near-side ridge yield and slightly overestimates the away-side yield for the interval 1~$<\it{p}_{\rm{T}}<$~2~GeV/$c$. However, the $\pythiashoving$ model underestimates the near-side ridge yield for $\it{p}_{\rm{T}}>$~2~GeV/$c$. The $\pythiam$ model does not show any near-side ridge as expected. It slightly underestimates the away-side peak for 1~$<\it{p}_{\rm{T}}<$~2~GeV/$c$ and provides good estimates for $\it{p}_{\rm{T}}>$~2~GeV/$c$. On the other hand, $\epos$ describes the shape of the ridge yield quantitatively better in the 2~$<\pttrigassoc<$~4~GeV/$c$ range, while overestimating the near-side ridge yield for $\pttrigassoc<$~2~GeV/$c$ range.

\begin{figure}[h!]
	\centering
	\subfigure{\includegraphics[width=0.89\textwidth]{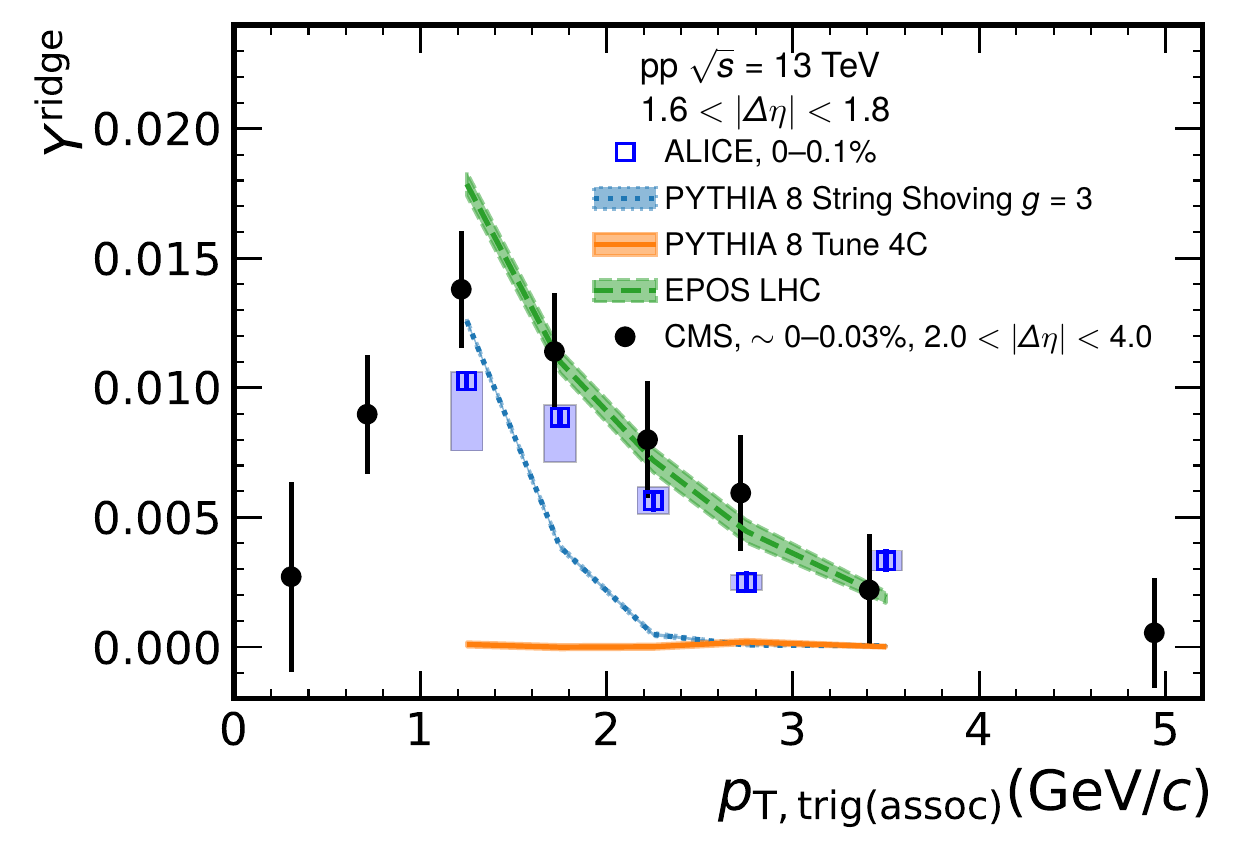}}
	\caption{ Fully corrected near-side ridge yield as a function transverse momentum. The open blue boxes denote the measurement by ALICE. The statistical and systematic uncertainties are shown as vertical bars and boxes, respectively. The CMS measurement~\cite{Khachatryan:2015lva} is represented by filled circles and extends down to lower $\pt$ due to the larger $\Delta\eta$ acceptance (see Sec.~\ref{sec:ana}). The three lines show model predictions from $\pythiam$ (blue dotted line), $\pythiashoving$ (orange line), and $\epos$ (green dashed line).}
	\label{fig:PlotYSpect}
\end{figure}

Figure~\ref{fig:PlotYSpect} shows the near-side ridge yield measured in high-multiplicity events as a function of $\pttrigassoc$. The measurement is compared with the result from CMS~\cite{Khachatryan:2015lva}.
Considering the differences in acceptance and the chosen multiplicity estimator of both measurements, perfect agreement between the two sets of results is not expected. The measurement is also compared with model calculations. As expected, the PYTHIA~8 model with Tune 4C does not produce a near-side ridge because it is not designed to account for this effect. The $\pythiashoving$ model describes the yield qualitatively, however the predicted yield decreases more rapidly than the measured one for increasing $\pttrigassoc$. The $\epos$ model, unlike the $\pythiashoving$ model, describes well the $\pt$ dependence of the ridge yield for the range $\it{p}_{\rm{T}}>$~2~GeV/$c$ , while predicting larger yields for $\it{p}_{\rm{T}}<$~2 GeV/$c$.

\subsection{Event-scale dependence of the ridge yield}
\begin{figure}[h!]
	\centering
	\subfigure{ \includegraphics[width=0.47 \textwidth]{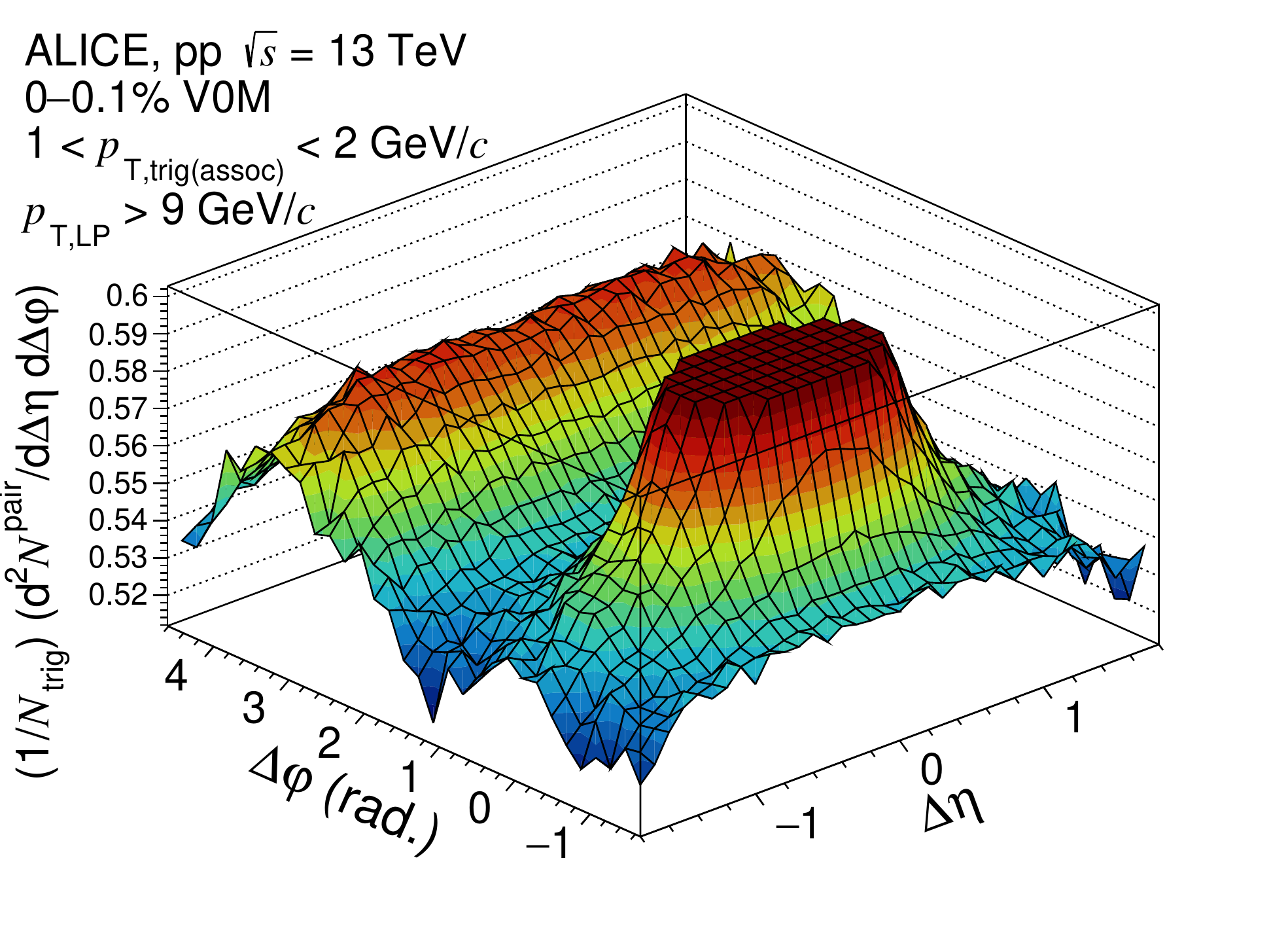} }
	\subfigure{ \includegraphics[width=0.47 \textwidth]{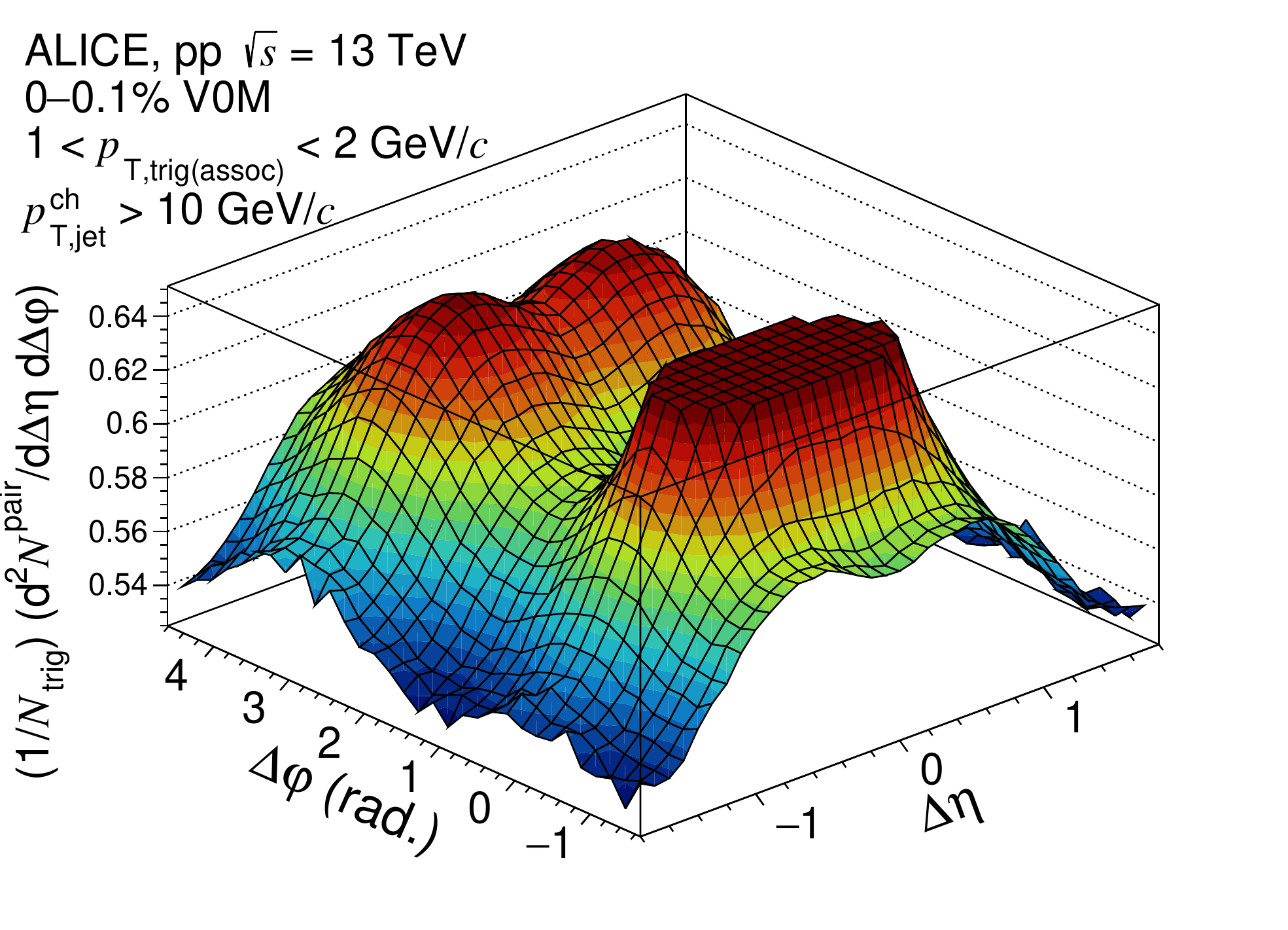} }
	\caption{ Two-dimensional correlation functions as a function of $\Delta\eta$ and $\Delta\varphi$ in high-multiplicity events including a selection on the event-scale. The interval of $\pttrig$ and $\ptassoc$ is 1~$<\pttrigassoc<$~2~GeV/$c$. Left: HM events with a $\ptlead>$~9~GeV/$c$ leading track. Right: HM events with a $\ptjet>$~10~GeV/$c$.}
	\label{fig:PlotCorrHMTSel}
\end{figure}

The ridge yield is further studied with respect to two different event-scales. In the first measurement, the event-scale is set by requiring a minimum $\pt$ on the leading particle in each event (denoted as $\ptlead$), while in the second measurement, a minimum $\pt$ is imposed on the leading jet (denoted as $\ptjet$). 

Figure~\ref{fig:PlotCorrHMTSel} shows that the ridge structure for 1~$< \pttrig\,(\ptassoc) <$~2 GeV/$c$ still persists in high-multiplicity pp collisions with $\ptlead>$~9~GeV/$c$ (left) and $\ptjet>$~10~GeV/$c$ (right).  It is worth noting that the correlation function obtained with the minimum $\ptjet$ selection has a double peak structure which is oriented along the $\Delta\eta$ axis at $\Delta\varphi=\pi$. This structure emerges due to the restricted acceptance of the jet tagging, $|\eta_{\rm{jet}}|<$~0.4.

\begin{figure}[h!]
	\centering
	\includegraphics[width=0.99\linewidth]{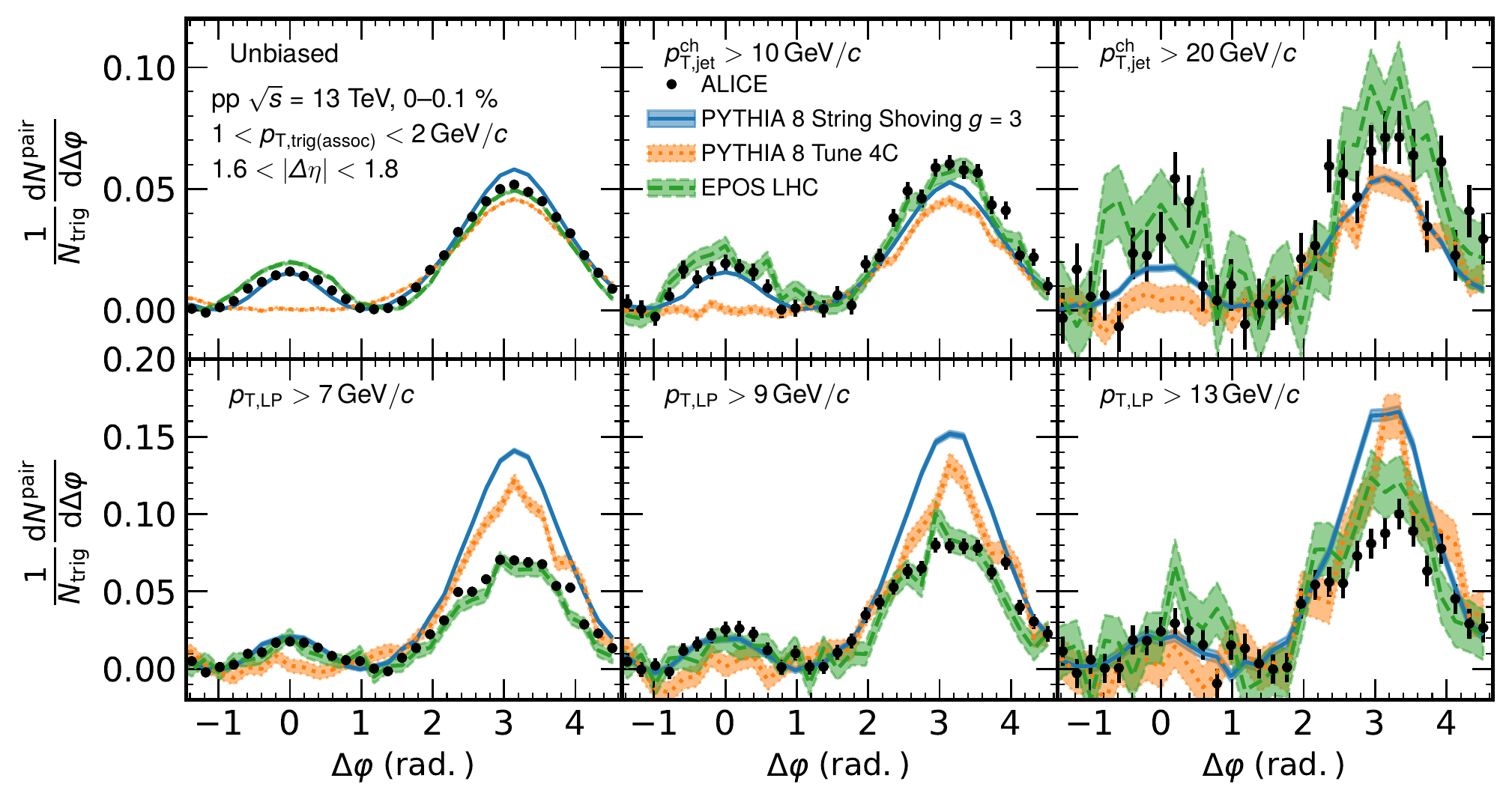}
	\caption{ One-dimensional $\Delta\varphi$ projections of the correlation functions constrained to 1.6~$<|\Delta\eta|<$~1.8 in HM events with an additional event-scale bias. Top: with an imposed selection on the leading jet $\pt$,  bottom: with an imposed selection on the leading particle $\pt$. ALICE data are compared with prediction of models.}
	\label{fig:PlotDeltaPhiESE}
\end{figure}

Figure~\ref{fig:PlotDeltaPhiESE} shows projected $\Delta\varphi$ distributions of the correlation functions in 1.6~$<|\Delta\eta|<$~1.8 with the minimum $\ptlead$ (lower) and $\ptjet$ (upper) requirement. Even with the event-scale selection, the ridge is still visible on the near-side. The near-side ridge peak increases as the thresholds of $\ptlead$ and $\ptjet$ increase compared to the one measured in unbiased events in Sec.~\ref{sec:resultunbiased}. The results are compared with $\pythiashoving$, $\pythiam$, and $\epos$ calculations. The near-side ridge peaks are qualitatively reproduced by $\pythiashoving$ and $\epos$ models. On the other hand, the $\pythiam$ does not show the  near-side ridge peak for neither of the two event-scale selections, but it gives compatible results for the away-side yield just like the $\pythiashoving$ model.

\begin{figure}[h!]
	\centering
	\includegraphics[width=0.89\linewidth]{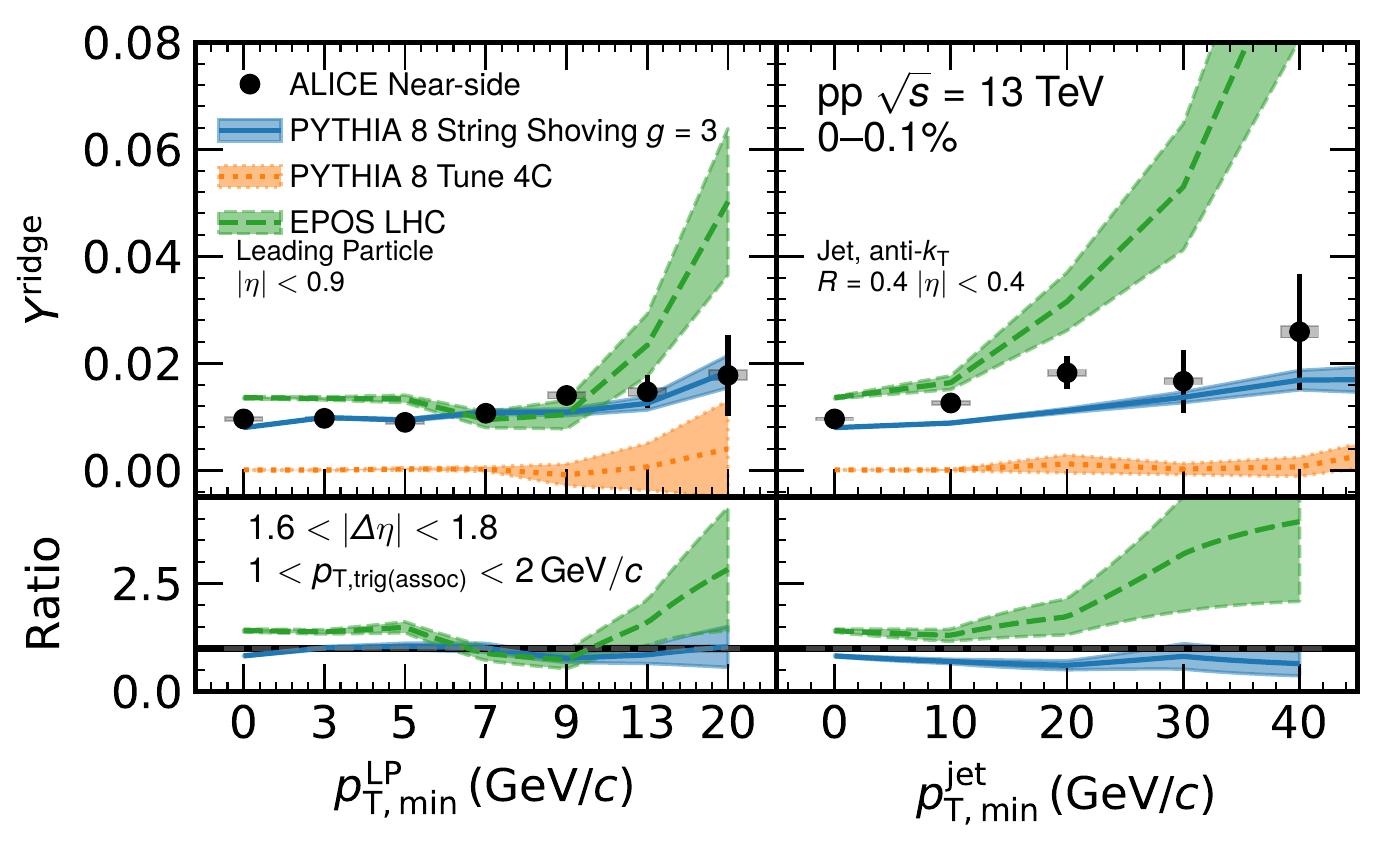}
	\caption{Near-side ridge yield as a function of the $\it{p}^{\rm{LP}}_{\rm{T,min}}$ (left) and $\it{p}^{\rm{Jet}}_{\rm{T,min}}$ (right). Data points (filled circles) show the ALICE measurement. The statistical and systematic uncertainties are shown as vertical bars and boxes, respectively. As the ridge yield is obtained in the same operational way for data and models, the upper limit of the systematic uncertainty due to jet contamination, which is 18.9\%, is not included in the figure. The data are compared with predictions of models which are represented by colored bands. The bottom panel shows a ratio of the models to the data. The uncertainty of the data is represented by the gray band centered around unity.}
	\label{fig:RidgeYield_ESE}
\end{figure}

The ridge yields as function of the minimum $\ptlead$ ($\it{p}^{\rm{LP}}_{\rm{T,min}}$) and $\ptjet$ ($\it{p}^{\rm{jet}}_{\rm{T,min}}$) selections are shown in Fig.~\ref{fig:RidgeYield_ESE}. High-multiplicity events with imposed event-scale bias exhibit increased ridge yields when compared to unbiased HM events. A small increase of the ridge yields as a function of $\ptlead$ or $\ptjet$ is observed and there is no difference between the two event-scale selections within the uncertainties. Comparisons to model calculations show that $\pythiashoving$ provides a comparable trend with data, but underestimates the ridge yield. On the other hand, $\epos$ overestimates the ridge yield while providing a trend comparable with the data. The origin of the enhanced ridge yields for higher momentum jet-tagged events is not clear to date but it might be related to the expected smaller impact parameters for dijet or multi-jet production events as studied in~\cite{Frankfurt:2010ea}.

\begin{figure}[h!]
	\centering
	\includegraphics[width=0.89\linewidth]{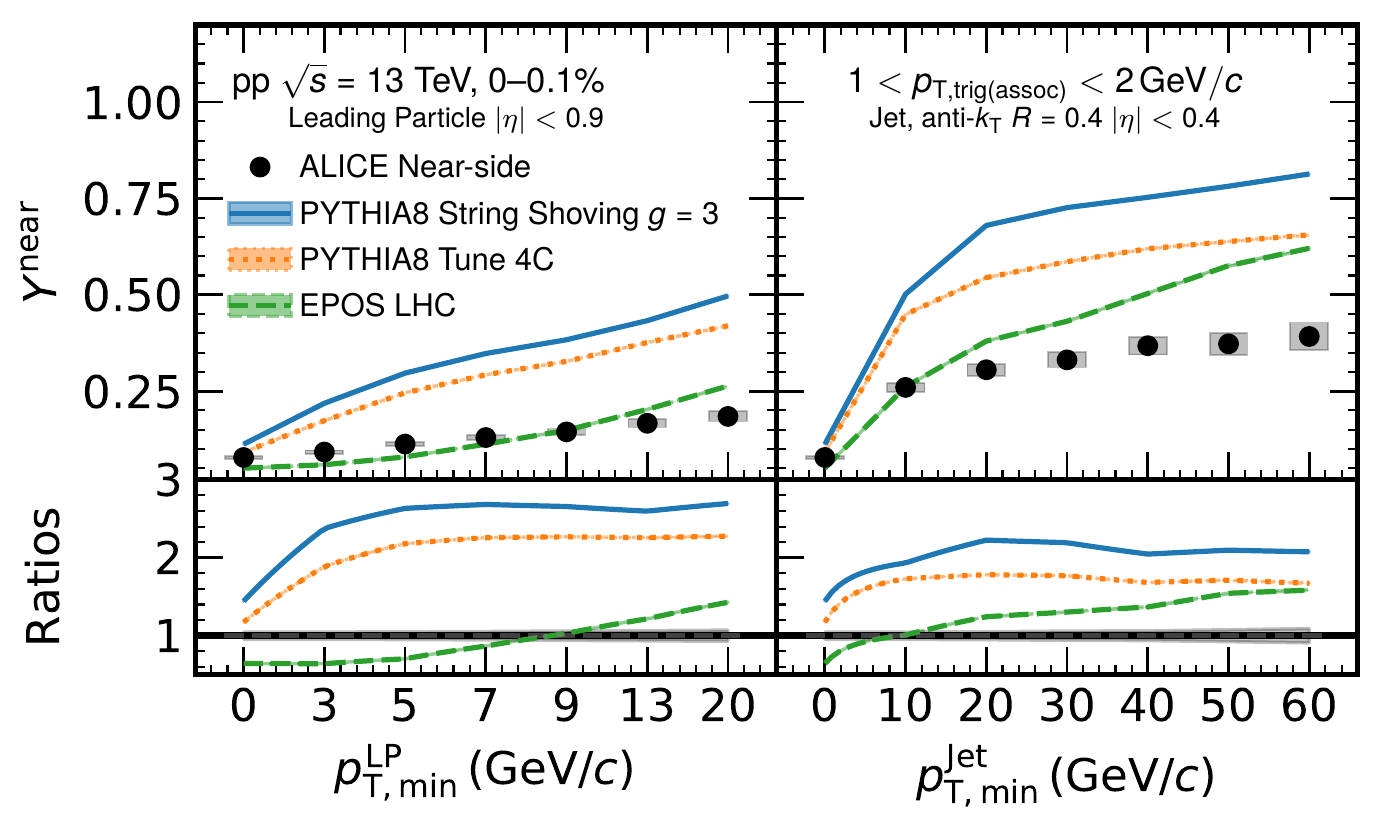}
	\caption{ Near-side jet-like peak yield as a function of the $\it{p}^{\rm{LP}}_{\rm{T,min}}$ (left) and $\it{p}^{\rm{jet}}_{\rm{T,min}}$ (right). The filled circles show measurement with ALICE. The statistical and systematic uncertainties are shown as vertical bars and boxes, respectively. The measurements are compared with model descriptions from $\pythiam$, $\pythiashoving$, and $\epos$ for both selections. The total uncertainty of the ratio is represented by the gray band centered around unity.}
	\label{fig:JetYield_ESE}
\end{figure}

Finally, the near-side jet-like peak yield is measured as a function of minimum $\ptlead$ and $\ptjet$ in Fig.~\ref{fig:JetYield_ESE} to further test the models that aim to describe the near-side ridge. $\epos$ provides comparable estimates of the near-side jet-like peak yield, while $\pythiam$ and $\pythiashoving$ overestimate the near-side yields for both event selections.

In all models if the ridge is due to final-state interactions, e.g., $\epos$ and PYTHIA~8 String Shoving, one also expects the near-side jet-like peak yield to be affected. This can be observed when comparing the measured near-side jet yields with PYTHIA~8 calculations with and without String Shoving. The new ALICE results therefore provide constraints beyond traditional ridge measurements that challenge existing models.


\section{Conclusions}
\label{sec:summary}

Long- and short-range correlations for pairs of charged particles with 1~$ < \pt < $~4~GeV/$c$ are studied in pp collisions at $\sqrt{s} = 13$~TeV with a focus on high-multiplicity events. The ridge and near-side jet yields are extracted and their event scale dependence have been studied. The obtained long-range ridge yields are compatible to those observed by the CMS Collaboration~\cite{Khachatryan:2015lva}.
The $\pythiashoving$ model describes the observed yields qualitatively but the yields it predicts decrease more rapidly with increasing $\pttrigassoc$ than those measured. On the other hand, the $\epos$ model gives a better description for the $\pttrigassoc$ dependence while overestimating the ridge yield for $\pttrigassoc<$~2 GeV/$c$. Finally, no long-range ridge is formed in the $\pythiam$ model.

The ridge yields are further studied in high-multiplicity events biased with additional event-scale selections, which impose a minimum transverse momentum cutoff on a leading track or jet. The ridge structure still persists with both selection criteria. The ridge yield increases as $\ptlead$ and $\ptjet$ increase. $\pythiashoving$ and $\epos$ estimate qualitatively the trends for the event-scale selections. However, the former underestimates and the latter overestimates it. The model predictions are also compared with the yield of the near-side jet-like correlation measured in the biased events. The evolution of the near-side jet yield as a function of event-scale $\pt$ is better captured by $\epos$, while the $\pythiashoving$ calculation tends to overshoot the data. 
The results might open a new way of studying the impact parameter dependence of small systems with jet tagged events in the future and will help to constrain the physical origins of long-range correlations.


\newenvironment{acknowledgement}{\relax}{\relax}
\begin{acknowledgement}
\section*{Acknowledgements}
\noindent The ALICE Collaboration would like to thank Christian Bierlich for providing the PYTHIA8 String Shoving configurations.

The ALICE Collaboration would like to thank all its engineers and technicians for their invaluable contributions to the construction of the experiment and the CERN accelerator teams for the outstanding performance of the LHC complex.
The ALICE Collaboration gratefully acknowledges the resources and support provided by all Grid centres and the Worldwide LHC Computing Grid (WLCG) collaboration.
The ALICE Collaboration acknowledges the following funding agencies for their support in building and running the ALICE detector:
A. I. Alikhanyan National Science Laboratory (Yerevan Physics Institute) Foundation (ANSL), State Committee of Science and World Federation of Scientists (WFS), Armenia;
Austrian Academy of Sciences, Austrian Science Fund (FWF): [M 2467-N36] and Nationalstiftung f\"{u}r Forschung, Technologie und Entwicklung, Austria;
Ministry of Communications and High Technologies, National Nuclear Research Center, Azerbaijan;
Conselho Nacional de Desenvolvimento Cient\'{\i}fico e Tecnol\'{o}gico (CNPq), Financiadora de Estudos e Projetos (Finep), Funda\c{c}\~{a}o de Amparo \`{a} Pesquisa do Estado de S\~{a}o Paulo (FAPESP) and Universidade Federal do Rio Grande do Sul (UFRGS), Brazil;
Ministry of Education of China (MOEC) , Ministry of Science \& Technology of China (MSTC) and National Natural Science Foundation of China (NSFC), China;
Ministry of Science and Education and Croatian Science Foundation, Croatia;
Centro de Aplicaciones Tecnol\'{o}gicas y Desarrollo Nuclear (CEADEN), Cubaenerg\'{\i}a, Cuba;
Ministry of Education, Youth and Sports of the Czech Republic, Czech Republic;
The Danish Council for Independent Research | Natural Sciences, the VILLUM FONDEN and Danish National Research Foundation (DNRF), Denmark;
Helsinki Institute of Physics (HIP), Finland;
Commissariat \`{a} l'Energie Atomique (CEA) and Institut National de Physique Nucl\'{e}aire et de Physique des Particules (IN2P3) and Centre National de la Recherche Scientifique (CNRS), France;
Bundesministerium f\"{u}r Bildung und Forschung (BMBF) and GSI Helmholtzzentrum f\"{u}r Schwerionenforschung GmbH, Germany;
General Secretariat for Research and Technology, Ministry of Education, Research and Religions, Greece;
National Research, Development and Innovation Office, Hungary;
Department of Atomic Energy Government of India (DAE), Department of Science and Technology, Government of India (DST), University Grants Commission, Government of India (UGC) and Council of Scientific and Industrial Research (CSIR), India;
Indonesian Institute of Science, Indonesia;
Istituto Nazionale di Fisica Nucleare (INFN), Italy;
Institute for Innovative Science and Technology , Nagasaki Institute of Applied Science (IIST), Japanese Ministry of Education, Culture, Sports, Science and Technology (MEXT) and Japan Society for the Promotion of Science (JSPS) KAKENHI, Japan;
Consejo Nacional de Ciencia (CONACYT) y Tecnolog\'{i}a, through Fondo de Cooperaci\'{o}n Internacional en Ciencia y Tecnolog\'{i}a (FONCICYT) and Direcci\'{o}n General de Asuntos del Personal Academico (DGAPA), Mexico;
Nederlandse Organisatie voor Wetenschappelijk Onderzoek (NWO), Netherlands;
The Research Council of Norway, Norway;
Commission on Science and Technology for Sustainable Development in the South (COMSATS), Pakistan;
Pontificia Universidad Cat\'{o}lica del Per\'{u}, Peru;
Ministry of Science and Higher Education, National Science Centre and WUT ID-UB, Poland;
Korea Institute of Science and Technology Information and National Research Foundation of Korea (NRF), Republic of Korea;
Ministry of Education and Scientific Research, Institute of Atomic Physics and Ministry of Research and Innovation and Institute of Atomic Physics, Romania;
Joint Institute for Nuclear Research (JINR), Ministry of Education and Science of the Russian Federation, National Research Centre Kurchatov Institute, Russian Science Foundation and Russian Foundation for Basic Research, Russia;
Ministry of Education, Science, Research and Sport of the Slovak Republic, Slovakia;
National Research Foundation of South Africa, South Africa;
Swedish Research Council (VR) and Knut \& Alice Wallenberg Foundation (KAW), Sweden;
European Organization for Nuclear Research, Switzerland;
Suranaree University of Technology (SUT), National Science and Technology Development Agency (NSDTA) and Office of the Higher Education Commission under NRU project of Thailand, Thailand;
Turkish Atomic Energy Agency (TAEK), Turkey;
National Academy of  Sciences of Ukraine, Ukraine;
Science and Technology Facilities Council (STFC), United Kingdom;
National Science Foundation of the United States of America (NSF) and United States Department of Energy, Office of Nuclear Physics (DOE NP), United States of America.
\end{acknowledgement}

\bibliographystyle{utphys}
\bibliography{paper.bib}

\newpage
\appendix

%
%

\section{The ALICE Collaboration}
\label{app:collab}
\begingroup
\small
\begin{flushleft} 


S.~Acharya$^{\rm 142}$, 
D.~Adamov\'{a}$^{\rm 97}$, 
A.~Adler$^{\rm 75}$, 
J.~Adolfsson$^{\rm 82}$, 
G.~Aglieri Rinella$^{\rm 35}$, 
M.~Agnello$^{\rm 31}$, 
N.~Agrawal$^{\rm 55}$, 
Z.~Ahammed$^{\rm 142}$, 
S.~Ahmad$^{\rm 16}$, 
S.U.~Ahn$^{\rm 77}$
% ...

\section*{Affiliation notes}

$^{\rm I}$ Deceased\\
$^{\rm II}$ Also at: Italian National Agency for New Technologies, Energy and Sustainable Economic Development (ENEA), Bologna, Italy\\
$^{\rm III}$ Also at: Dipartimento DET del Politecnico di Torino, Turin, Italy\\
$^{\rm IV}$ Also at: M.V. Lomonosov Moscow State University, D.V. Skobeltsyn Institute of Nuclear, Physics, Moscow, Russia\\
$^{\rm V}$ Also at: Institute of Theoretical Physics, University of Wroclaw, Poland\\

\section*{Collaboration Institutes}

$^{1}$ A.I. Alikhanyan National Science Laboratory (Yerevan Physics Institute) Foundation, Yerevan, Armenia\\
$^{2}$ AGH University of Science and Technology, Cracow, Poland\\
$^{3}$ Bogolyubov Institute for Theoretical Physics, National Academy of Sciences of Ukraine, Kiev, Ukraine\\
$^{4}$ Bose Institute, Department of Physics  and Centre for Astroparticle Physics and Space Science (CAPSS), Kolkata, India\\
$^{5}$ Budker Institute for Nuclear Physics, Novosibirsk, Russia\\
$^{6}$ California Polytechnic State University, San Luis Obispo, California, United States\\
$^{7}$ Central China Normal University, Wuhan, China\\
$^{8}$ Centro de Aplicaciones Tecnol\'{o}gicas y Desarrollo Nuclear (CEADEN), Havana, Cuba\\
$^{9}$ Centro de Investigaci\'{o}n y de Estudios Avanzados (CINVESTAV), Mexico City and M\'{e}rida, Mexico\\
$^{10}$ Chicago State University, Chicago, Illinois, United States\\
$^{11}$ China Institute of Atomic Energy, Beijing, China\\
$^{12}$ Chungbuk National University, Cheongju, Republic of Korea\\
$^{13}$ Comenius University Bratislava, Faculty of Mathematics, Physics and Informatics, Bratislava, Slovakia\\
$^{14}$ COMSATS University Islamabad, Islamabad, Pakistan\\
$^{15}$ Creighton University, Omaha, Nebraska, United States\\
$^{16}$ Department of Physics, Aligarh Muslim University, Aligarh, India\\
$^{17}$ Department of Physics, Pusan National University, Pusan, Republic of Korea\\
$^{18}$ Department of Physics, Sejong University, Seoul, Republic of Korea\\
$^{19}$ Department of Physics, University of California, Berkeley, California, United States\\
$^{20}$ Department of Physics, University of Oslo, Oslo, Norway\\
$^{21}$ Department of Physics and Technology, University of Bergen, Bergen, Norway\\
$^{22}$ Dipartimento di Fisica dell'Universit\`{a} 'La Sapienza' and Sezione INFN, Rome, Italy\\
$^{23}$ Dipartimento di Fisica dell'Universit\`{a} and Sezione INFN, Cagliari, Italy\\
$^{24}$ Dipartimento di Fisica dell'Universit\`{a} and Sezione INFN, Trieste, Italy\\
$^{25}$ Dipartimento di Fisica dell'Universit\`{a} and Sezione INFN, Turin, Italy\\
$^{26}$ Dipartimento di Fisica e Astronomia dell'Universit\`{a} and Sezione INFN, Bologna, Italy\\
$^{27}$ Dipartimento di Fisica e Astronomia dell'Universit\`{a} and Sezione INFN, Catania, Italy\\
$^{28}$ Dipartimento di Fisica e Astronomia dell'Universit\`{a} and Sezione INFN, Padova, Italy\\
$^{29}$ Dipartimento di Fisica e Nucleare e Teorica, Universit\`{a} di Pavia  and Sezione INFN, Pavia, Italy\\
$^{30}$ Dipartimento di Fisica `E.R.~Caianiello' dell'Universit\`{a} and Gruppo Collegato INFN, Salerno, Italy\\
$^{31}$ Dipartimento DISAT del Politecnico and Sezione INFN, Turin, Italy\\
$^{32}$ Dipartimento di Scienze e Innovazione Tecnologica dell'Universit\`{a} del Piemonte Orientale and INFN Sezione di Torino, Alessandria, Italy\\
$^{33}$ Dipartimento di Scienze MIFT, Universit\`{a} di Messina, Messina, Italy\\
$^{34}$ Dipartimento Interateneo di Fisica `M.~Merlin' and Sezione INFN, Bari, Italy\\
$^{35}$ European Organization for Nuclear Research (CERN), Geneva, Switzerland\\
$^{36}$ Faculty of Electrical Engineering, Mechanical Engineering and Naval Architecture, University of Split, Split, Croatia\\
$^{37}$ Faculty of Engineering and Science, Western Norway University of Applied Sciences, Bergen, Norway\\
$^{38}$ Faculty of Nuclear Sciences and Physical Engineering, Czech Technical University in Prague, Prague, Czech Republic\\
$^{39}$ Faculty of Science, P.J.~\v{S}af\'{a}rik University, Ko\v{s}ice, Slovakia\\
$^{40}$ Frankfurt Institute for Advanced Studies, Johann Wolfgang Goethe-Universit\"{a}t Frankfurt, Frankfurt, Germany\\
$^{41}$ Fudan University, Shanghai, China\\
$^{42}$ Gangneung-Wonju National University, Gangneung, Republic of Korea\\
$^{43}$ Gauhati University, Department of Physics, Guwahati, India\\
$^{44}$ Helmholtz-Institut f\"{u}r Strahlen- und Kernphysik, Rheinische Friedrich-Wilhelms-Universit\"{a}t Bonn, Bonn, Germany\\
$^{45}$ Helsinki Institute of Physics (HIP), Helsinki, Finland\\
$^{46}$ High Energy Physics Group,  Universidad Aut\'{o}noma de Puebla, Puebla, Mexico\\
$^{47}$ Hiroshima University, Hiroshima, Japan\\
$^{48}$ Hochschule Worms, Zentrum  f\"{u}r Technologietransfer und Telekommunikation (ZTT), Worms, Germany\\
$^{49}$ Horia Hulubei National Institute of Physics and Nuclear Engineering, Bucharest, Romania\\
$^{50}$ Indian Institute of Technology Bombay (IIT), Mumbai, India\\
$^{51}$ Indian Institute of Technology Indore, Indore, India\\
$^{52}$ Indonesian Institute of Sciences, Jakarta, Indonesia\\
$^{53}$ INFN, Laboratori Nazionali di Frascati, Frascati, Italy\\
$^{54}$ INFN, Sezione di Bari, Bari, Italy\\
$^{55}$ INFN, Sezione di Bologna, Bologna, Italy\\
$^{56}$ INFN, Sezione di Cagliari, Cagliari, Italy\\
$^{57}$ INFN, Sezione di Catania, Catania, Italy\\
$^{58}$ INFN, Sezione di Padova, Padova, Italy\\
$^{59}$ INFN, Sezione di Roma, Rome, Italy\\
$^{60}$ INFN, Sezione di Torino, Turin, Italy\\
$^{61}$ INFN, Sezione di Trieste, Trieste, Italy\\
$^{62}$ Inha University, Incheon, Republic of Korea\\
$^{63}$ Institute for Gravitational and Subatomic Physics (GRASP), Utrecht University/Nikhef, Utrecht, Netherlands\\
$^{64}$ Institute for Nuclear Research, Academy of Sciences, Moscow, Russia\\
$^{65}$ Institute of Experimental Physics, Slovak Academy of Sciences, Ko\v{s}ice, Slovakia\\
$^{66}$ Institute of Physics, Homi Bhabha National Institute, Bhubaneswar, India\\
$^{67}$ Institute of Physics of the Czech Academy of Sciences, Prague, Czech Republic\\
$^{68}$ Institute of Space Science (ISS), Bucharest, Romania\\
$^{69}$ Institut f\"{u}r Kernphysik, Johann Wolfgang Goethe-Universit\"{a}t Frankfurt, Frankfurt, Germany\\
$^{70}$ Instituto de Ciencias Nucleares, Universidad Nacional Aut\'{o}noma de M\'{e}xico, Mexico City, Mexico\\
$^{71}$ Instituto de F\'{i}sica, Universidade Federal do Rio Grande do Sul (UFRGS), Porto Alegre, Brazil\\
$^{72}$ Instituto de F\'{\i}sica, Universidad Nacional Aut\'{o}noma de M\'{e}xico, Mexico City, Mexico\\
$^{73}$ iThemba LABS, National Research Foundation, Somerset West, South Africa\\
$^{74}$ Jeonbuk National University, Jeonju, Republic of Korea\\
$^{75}$ Johann-Wolfgang-Goethe Universit\"{a}t Frankfurt Institut f\"{u}r Informatik, Fachbereich Informatik und Mathematik, Frankfurt, Germany\\
$^{76}$ Joint Institute for Nuclear Research (JINR), Dubna, Russia\\
$^{77}$ Korea Institute of Science and Technology Information, Daejeon, Republic of Korea\\
$^{78}$ KTO Karatay University, Konya, Turkey\\
$^{79}$ Laboratoire de Physique des 2 Infinis, Ir\`{e}ne Joliot-Curie, Orsay, France\\
$^{80}$ Laboratoire de Physique Subatomique et de Cosmologie, Universit\'{e} Grenoble-Alpes, CNRS-IN2P3, Grenoble, France\\
$^{81}$ Lawrence Berkeley National Laboratory, Berkeley, California, United States\\
$^{82}$ Lund University Department of Physics, Division of Particle Physics, Lund, Sweden\\
$^{83}$ Moscow Institute for Physics and Technology, Moscow, Russia\\
$^{84}$ Nagasaki Institute of Applied Science, Nagasaki, Japan\\
$^{85}$ Nara Women{'}s University (NWU), Nara, Japan\\
$^{86}$ National and Kapodistrian University of Athens, School of Science, Department of Physics , Athens, Greece\\
$^{87}$ National Centre for Nuclear Research, Warsaw, Poland\\
$^{88}$ National Institute of Science Education and Research, Homi Bhabha National Institute, Jatni, India\\
$^{89}$ National Nuclear Research Center, Baku, Azerbaijan\\
$^{90}$ National Research Centre Kurchatov Institute, Moscow, Russia\\
$^{91}$ Niels Bohr Institute, University of Copenhagen, Copenhagen, Denmark\\
$^{92}$ Nikhef, National institute for subatomic physics, Amsterdam, Netherlands\\
$^{93}$ NRC Kurchatov Institute IHEP, Protvino, Russia\\
$^{94}$ NRC \guillemotleft Kurchatov\guillemotright  Institute - ITEP, Moscow, Russia\\
$^{95}$ NRNU Moscow Engineering Physics Institute, Moscow, Russia\\
$^{96}$ Nuclear Physics Group, STFC Daresbury Laboratory, Daresbury, United Kingdom\\
$^{97}$ Nuclear Physics Institute of the Czech Academy of Sciences, \v{R}e\v{z} u Prahy, Czech Republic\\
$^{98}$ Oak Ridge National Laboratory, Oak Ridge, Tennessee, United States\\
$^{99}$ Ohio State University, Columbus, Ohio, United States\\
$^{100}$ Petersburg Nuclear Physics Institute, Gatchina, Russia\\
$^{101}$ Physics department, Faculty of science, University of Zagreb, Zagreb, Croatia\\
$^{102}$ Physics Department, Panjab University, Chandigarh, India\\
$^{103}$ Physics Department, University of Jammu, Jammu, India\\
$^{104}$ Physics Department, University of Rajasthan, Jaipur, India\\
$^{105}$ Physikalisches Institut, Eberhard-Karls-Universit\"{a}t T\"{u}bingen, T\"{u}bingen, Germany\\
$^{106}$ Physikalisches Institut, Ruprecht-Karls-Universit\"{a}t Heidelberg, Heidelberg, Germany\\
$^{107}$ Physik Department, Technische Universit\"{a}t M\"{u}nchen, Munich, Germany\\
$^{108}$ Politecnico di Bari and Sezione INFN, Bari, Italy\\
$^{109}$ Research Division and ExtreMe Matter Institute EMMI, GSI Helmholtzzentrum f\"ur Schwerionenforschung GmbH, Darmstadt, Germany\\
$^{110}$ Rudjer Bo\v{s}kovi\'{c} Institute, Zagreb, Croatia\\
$^{111}$ Russian Federal Nuclear Center (VNIIEF), Sarov, Russia\\
$^{112}$ Saha Institute of Nuclear Physics, Homi Bhabha National Institute, Kolkata, India\\
$^{113}$ School of Physics and Astronomy, University of Birmingham, Birmingham, United Kingdom\\
$^{114}$ Secci\'{o}n F\'{\i}sica, Departamento de Ciencias, Pontificia Universidad Cat\'{o}lica del Per\'{u}, Lima, Peru\\
$^{115}$ St. Petersburg State University, St. Petersburg, Russia\\
$^{116}$ Stefan Meyer Institut f\"{u}r Subatomare Physik (SMI), Vienna, Austria\\
$^{117}$ SUBATECH, IMT Atlantique, Universit\'{e} de Nantes, CNRS-IN2P3, Nantes, France\\
$^{118}$ Suranaree University of Technology, Nakhon Ratchasima, Thailand\\
$^{119}$ Technical University of Ko\v{s}ice, Ko\v{s}ice, Slovakia\\
$^{120}$ The Henryk Niewodniczanski Institute of Nuclear Physics, Polish Academy of Sciences, Cracow, Poland\\
$^{121}$ The University of Texas at Austin, Austin, Texas, United States\\
$^{122}$ Universidad Aut\'{o}noma de Sinaloa, Culiac\'{a}n, Mexico\\
$^{123}$ Universidade de S\~{a}o Paulo (USP), S\~{a}o Paulo, Brazil\\
$^{124}$ Universidade Estadual de Campinas (UNICAMP), Campinas, Brazil\\
$^{125}$ Universidade Federal do ABC, Santo Andre, Brazil\\
$^{126}$ University of Cape Town, Cape Town, South Africa\\
$^{127}$ University of Houston, Houston, Texas, United States\\
$^{128}$ University of Jyv\"{a}skyl\"{a}, Jyv\"{a}skyl\"{a}, Finland\\
$^{129}$ University of Liverpool, Liverpool, United Kingdom\\
$^{130}$ University of Science and Technology of China, Hefei, China\\
$^{131}$ University of South-Eastern Norway, Tonsberg, Norway\\
$^{132}$ University of Tennessee, Knoxville, Tennessee, United States\\
$^{133}$ University of the Witwatersrand, Johannesburg, South Africa\\
$^{134}$ University of Tokyo, Tokyo, Japan\\
$^{135}$ University of Tsukuba, Tsukuba, Japan\\
$^{136}$ Universit\'{e} Clermont Auvergne, CNRS/IN2P3, LPC, Clermont-Ferrand, France\\
$^{137}$ Universit\'{e} de Lyon, CNRS/IN2P3, Institut de Physique des 2 Infinis de Lyon , Lyon, France\\
$^{138}$ Universit\'{e} de Strasbourg, CNRS, IPHC UMR 7178, F-67000 Strasbourg, France, Strasbourg, France\\
$^{139}$ Universit\'{e} Paris-Saclay Centre d'Etudes de Saclay (CEA), IRFU, D\'{e}partment de Physique Nucl\'{e}aire (DPhN), Saclay, France\\
$^{140}$ Universit\`{a} degli Studi di Foggia, Foggia, Italy\\
$^{141}$ Universit\`{a} di Brescia and Sezione INFN, Brescia, Italy\\
$^{142}$ Variable Energy Cyclotron Centre, Homi Bhabha National Institute, Kolkata, India\\
$^{143}$ Warsaw University of Technology, Warsaw, Poland\\
$^{144}$ Wayne State University, Detroit, Michigan, United States\\
$^{145}$ Westf\"{a}lische Wilhelms-Universit\"{a}t M\"{u}nster, Institut f\"{u}r Kernphysik, M\"{u}nster, Germany\\
$^{146}$ Wigner Research Centre for Physics, Budapest, Hungary\\
$^{147}$ Yale University, New Haven, Connecticut, United States\\
$^{148}$ Yonsei University, Seoul, Republic of Korea\\

\bigskip 

\end{flushleft} 
\endgroup
\end{document}